\documentclass[aps,pre,twocolumn,superscriptaddress,showpacs,showkeys ]{revtex4-1}

\usepackage{graphicx}% Include figure files
\usepackage{dcolumn}% Align table columns on decimal point
\usepackage{bm}% bold math
\RequirePackage[colorlinks=true, allcolors=blue]{hyperref}

\graphicspath{{graphics/}}    % the path of image folder 
\begin{document}

% \title{Manuscript Title:\\with Forced Linebreak}% Force line breaks with \\
\title{Go viral or go broadcast? Characterizing the virality and growth of cascades} % Force line breaks with \\
% \title{Viral vs. broadcast: Characterizing the virality and growth of cascades} %

\author{Yafei Zhang}
\affiliation{Department of Automation, Shanghai Jiao Tong University, Shanghai 200240, China}
\affiliation{Department of Media and Communication; School of Data Science, City University of Hong Kong, Hong Kong S.A.R., China}

\author{Lin Wang}
\affiliation{Department of Automation, Shanghai Jiao Tong University, Shanghai 200240, China}

\author{Jonathan J. H. Zhu}
\thanks{Corresponding authors}
\affiliation{Department of Media and Communication; School of Data Science, City University of Hong Kong, Hong Kong S.A.R., China}

\author{Xiaofan Wang}
\thanks{Corresponding authors}
\affiliation{Department of Automation, Shanghai Jiao Tong University, Shanghai 200240, China}
\affiliation{Department of Automation, Shanghai University, Shanghai 200444, China}


\begin{abstract}
% An article usually includes an abstract, a concise summary of the work 
% covered at length in the main body of the article. 
Quantifying the virality of cascades is an important question across disciplines
such as 
the transmission of disease, the spread of information and the diffusion of innovations.
An appropriate virality metric should be able to disambiguate between
a shallow, broadcast-like diffusion process and a deep, multi-generational branching process.
Although several valuable works
have been dedicated to this field,
most of them fail to take the position of the diffusion source 
into consideration,
which makes them fall into the trap of graph isomorphism and
would result in imprecise
estimation of 
cascade virality inevitably under certain circumstances.

In this paper, we propose a root-aware approach to {quantifying} the virality of cascades
with proper consideration of the root node in a diffusion tree. 
With applications on synthetic and empirical cascades,
we show the 
% property 
{properties}
and potential utility of the proposed virality measure.
Based on preferential attachment {mechanisms},
we further introduce a model to mimic the growth of cascades.
The proposed model
enables the interpolation between broadcast and viral spreading during the growth of cascades.
Through numerical simulations, we demonstrate the effectiveness of the proposed model in
characterizing
the virality of growing cascades.
Our work contributes to 
% strengthens 
the understanding of cascade virality and growth,
and could offer practical implications
in a range of policy domains
including viral marketing, infectious disease and information diffusion.

\end{abstract}

\maketitle

%\tableofcontents

\section{\label{sec:intro}Introduction}

Spreading is a ubiquitous process across disciplines.
Conceptually,
many spreading processes
can be viewed as tree-like cascades,
including
the retweeting or resharing of news in online social networks \cite{goel2015structural, vosoughi2018spread, cheng2014can, liben2008tracing, liang2018broadcast},
the generated discussion threads in online boards \cite{kumar2010dynamics, gomez2011modeling, medvedev2018modelling},
the outbreak or transmission of diseases \cite{pastor2015epidemic, ferreira2002reaction, hadfield2018nextstrain, moinet2018effect},
and
the diffusion of new products or services \cite{anderson2015global, banerjee2013diffusion, iribarren2011branching, zhang2015breadth}.
For instance, 
in a discussion thread, comments and the reply-to actions between them can be represented as nodes and edges in a rooted tree with the original post acting as the root \cite{medvedev2018modelling}.

In some cases, a single node could directly account for a large proportion of the whole diffusion process, which is a typical characteristic in the media broadcast industry;
while in others, diffusion is more likely to be driven by word-of-mouth 
% mechanism and circulates 
{mechanisms and circulate}
in a viral way, where each node only accounts for a small fraction of the whole diffusion process \cite{goel2015structural}.
For example,
one NBA Christmas game is able to attract millions of views through merely broadcast, but one tweet could reach millions as well through viral spreading.

\begin{figure}[!b]
\centering
\includegraphics[scale=1]{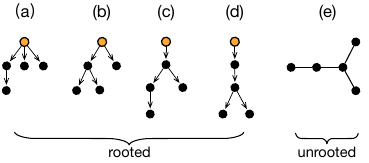}
\caption{\label{fig:cas_structure_sample} 
Examples of rooted cascades and the corresponding unrooted cascade.
The root node is colored in orange in each rooted cascade.
}
\end{figure}

As the backbone of so many diffusion processes, cascade structure
yields a valuable source to learn 
the extent to which a cascade grows in a broadcast (or breadth-first) or viral (or depth-first) way.
From the view of tree traversal,
we can simply interpret broadcast cascade as a result of ``breadth first search"
and viral cascade as a result of ``depth first search".
In this regard,
there have been several important attempts to quantify the 
virality of cascades  \cite{goel2015structural, cheng2014can, anderson2015global, vosoughi2018spread}, where the more viral a cascade is, the larger this virality quantity would be.
However, 
unlike many other undirected graphs,
once the root of a cascade is identified, the direction of its growth or flow is thus determined: from the root node to its descendants (see Fig. \ref{fig:cas_structure_sample}, (a)-(d)).
In other words, rooted cascades are actually 
directed graphs, rather than undirected graphs.

\begin{figure}[!tb]
\centering
\includegraphics[scale=0.8]{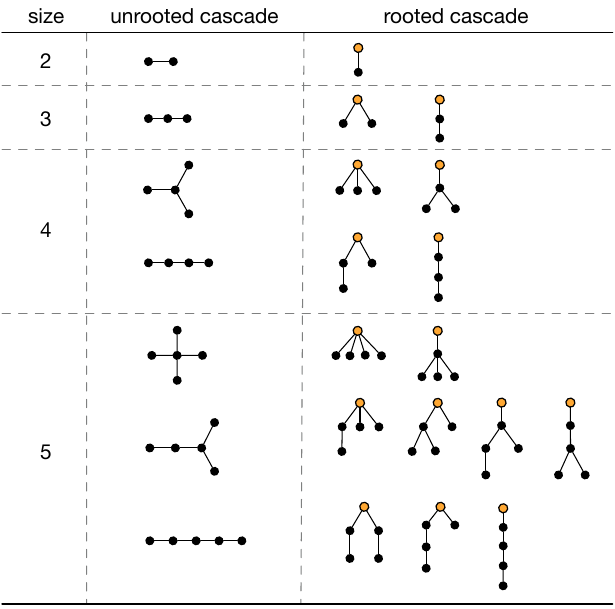}
\caption{\label{fig:cas_structure}
Unrooted cascades versus the corresponding rooted cascades.
For illustration {purposes}, only cascades of size less than or equal to 5 are shown.
The diffusion source is colored in orange in each rooted cascade.
For ease of visualization, the directed ties in each rooted cascade are not shown.
}
\end{figure}

For example, rooted cascades in Fig. \ref{fig:cas_structure_sample} (a)-(d) exhibit distinct structures, but if the root nodes are overlooked, they are deemed as the same structure under graph isomorphism (Fig. \ref{fig:cas_structure_sample} (e)).
More such examples are illustrated in Fig. \ref{fig:cas_structure}.
For ease of visualization, the directed ties in each rooted cascade are not shown hereafter, as the existence of the root node already implies the direction of the flow.
Therefore,
it's anticipated that
ignoring the root node 
and simply regarding cascades as undirected trees
% ignoring the root or the direction of a cascade 
would inevitably result in imprecise estimation of the virality of cascades under certain situations.

To remedy this, a root-aware approach is proposed to quantify the virality of cascades in this paper.
Specifically,
instead of computing the average distance between all pairs
of nodes in a cascade tree 
% (which is the case in Ref. \cite{goel2015structural}), 
(which is also termed
\textit{Wiener index} or \textit{structural virality} \cite{goel2015structural}),
we calculate 
the average distance between a node and its descendants,
and summing the computed distances over all nodes yields the amended virality of a cascade.
The proposed approach can operate in a recursive manner, and 
is able to provide {a} high-resolution 
% characterization 
depiction
of cascade virality.
% which would go unnoticed otherwise.
% 
Leveraging synthetic and empirical cascades, we show the property and potential utility
of the quantified cascade virality.

In light of preferential attachment {mechanisms} \cite{barabasi1999emergence, krapivsky2000connectivity, krapivsky2001organization, zhao2015seismic, medvedev2018modelling, krohn2019modelling, overgoor2019choosing}, this work further presents 
a 
% genealogy-based
cascade growth model
based on node genealogy
instead of merely node degree.
Specifically, the genealogy of a given node is defined as the subtree rooted at this node and contains only the given node itself and its descendants, and the probability that a new node attached to an existing node depends on the genealogy size of the target node at different generations.
Through numerical simulations, we further demonstrate the benefits brought by genealogy-based modeling (compared with degree-based modeling)
in capturing the virality landscape of cascades.

% The rest of the paper is organized as follows.
% In section 2, we...
% In section 3, we...

\begin{figure}[t]
\includegraphics[width=\hsize]{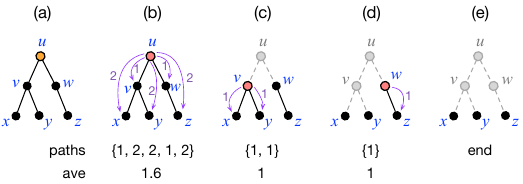}
\caption{\label{fig:cascade_alg} 
Algorithm for quantifying cascade virality.
To obtain the virality $V_T$ of a tree $T$, we imagine the tree is rooted at node $u$. Once node $u$ is removed, the tree would split into multiple subtrees (two subtrees in the given example), each rooted at one of $u$'s children (which are $v$ and $w$ in the case).
Repeat the root node removal process step by step, until no more nodes have any descendants.
Count the average path length between root nodes (which are $u$, $v$ and $w$ in the case) and their descendants, then add them up immediately yields the desired $V_T$.
% {\verb paths} indicates 
}
\end{figure}

\section{\label{sec:cas_virality}Cascade virality}

\subsection{Algorithm}

Quantifying the virality of cascades is a challenging task 
as it 
synthetically considers 
% involves
not only the size
but also the depth and branches of a given cascade tree.
Goel et al.
\cite{goel2015structural}
address this issue {using the} \textit{Wiener index},
which quantifies the average distance between any two nodes in a cascade tree,
as the desired virality measure which they term \textit{structural virality}.
This attempt
makes important advancement in quantitatively depicting the extent to which a cascade 
% grows in a viral way,
is viral, 
but it
fails to 
account for the root of a cascade
and inherently treats the cascade as an unrooted tree,
thereby hindering its ability to
differentiate cascades 
which are 
structurally different but will be considered as the same if the root nodes are neglected.
(see Fig. \ref{fig:cas_structure_sample} and Fig. \ref{fig:cas_structure}).
% 
% 
% We remedy this via root-aware mechanism.
To remedy this and more precisely quantify the virality of cascades,
we propose a root-aware approach.
The obtained virality score is then named as \textit{cascade virality}.

% In practice,
Specifically,
the cascade virality of a rooted tree is quantified by
the sum of the average path length between nodes and their descendants.
Formally,
for a rooted cascade tree $T$ of size $N$ $(N \geq 2)$,
the cascade virality of $T$ is 
\begin{equation}
V_T=\sum_{i=1}^{N}\bar{d}_{i},
\end{equation}
where
\begin{equation}
\bar{d}_{i}=\sum_{j\in \mathcal{D}_i} d_{ij}/|\mathcal{D}_i|,
\end{equation}
$d_{ij}$ denotes the distance from node $i$ to one of its descendant $j$,
$\mathcal{D}_i$ is the set of descendants of node $i$
and {$|\bullet|$} represents the size of a set.

A schematic of the above process is shown in  Fig. \ref{fig:cascade_alg}.
Starting from the root node $u$, the average distance from node $u$ to its descendants is counted, denoted as $\bar{d}_{u}=ave(1,2,2,1,2)=1.6$ (Fig. \ref{fig:cascade_alg}(b));
Suppose that node $u$ is then removed, thus resulting in two subtrees with 
node $v$ and $w$
(i.e., child nodes of node $u$)
act as new root {nodes} for each subtree;
The average distance from node $v$ and $w$ to their descendants is then computed as
$\bar{d}_{v}=ave(1,1)=1$ (Fig. \ref{fig:cascade_alg}(c)) and
$\bar{d}_{w}=ave(1)=1$ (Fig. \ref{fig:cascade_alg}(d));
Repeat the above node removal and average path length calculation process,
until no more nodes have {descendants} (Fig. \ref{fig:cascade_alg}(e)).
The cascade virality of $T$, $V_T$, is finally obtained by the
sum over all the calculated average path length $\bar{d}$,
which is $\bar{d}_{u}+\bar{d}_{v}+\bar{d}_{w}=3.6$ in the given example (Fig. \ref{fig:cascade_alg}).
Also note that the proposed approach to {quantifying} the cascade virality of a rooted tree is easy to implement in a recursive manner, which enables the time complexity in the actual execution.
A Python implementation of the algorithm is freely available online\footnote{\url{https://github.com/yflyzhang/cascade_virality}}.

\begin{figure}[tb]
\centering
\includegraphics[width=\hsize]{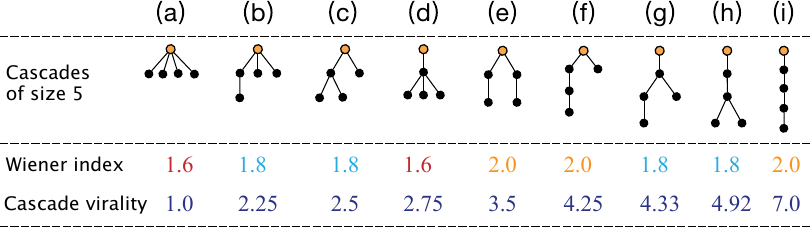}
\caption{\label{fig:viral_exs}
{
Cascades with five nodes and the corresponding Wiener index and cascade virality scores (shown below each cascade). Root nodes in each cascade are colored in orange.
For Wiener index, equal values are shown in the same color.
}
}
\end{figure}

\begin{figure*}[tb]
\centering
\includegraphics[width=16cm]{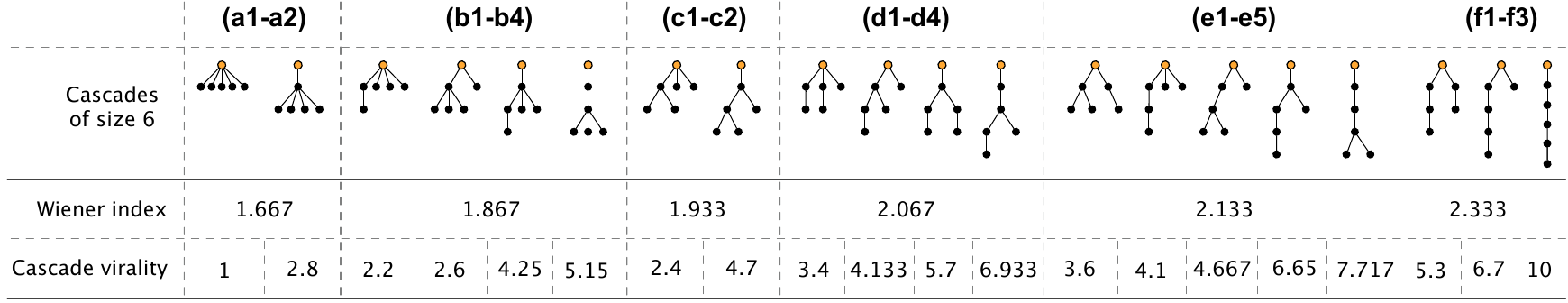}% Here is how to import EPS art
\caption{\label{fig:viral_exs_6}
{Cascades with six nodes and the corresponding Wiener index and cascade virality scores.}
}
\end{figure*}

{
Fig. \ref{fig:viral_exs} presents direct comparisons between Wiener index and cascade virality for cascades with five nodes.
As we can see from the figure, Wiener index, which quantifies the average distance between all pairs of nodes in a cascade tree, fails to distinguish several cascades as this virality measure doesn't take the root nodes into account.
For instance, cascades in Fig. \ref{fig:viral_exs}(e), \ref{fig:viral_exs}(f) and \ref{fig:viral_exs}(i) hold the same value in the context of Wiener index.
Similar conditions can also be found for cascades in Fig. \ref{fig:viral_exs}(a) and \ref{fig:viral_exs}(d) as well as cascades in Fig. \ref{fig:viral_exs}(b), \ref{fig:viral_exs}(c), \ref{fig:viral_exs}(g) and \ref{fig:viral_exs}(h).
In contrast, cascade virality is able to distinguish all the nine cascades in this scenario (last row in Fig. \ref{fig:viral_exs}).
Fig. \ref{fig:viral_exs_6} further illustrates cascades with six nodes and the corresponding values quantified by Wiener index and cascade virality.
Compared with Fig. \ref{fig:viral_exs}, Fig. \ref{fig:viral_exs_6} demonstrates that 
there are even more cascades that 
Wiener index fails to distinguish from each other.
Specifically, for the twenty cascades illustrated in Fig. \ref{fig:viral_exs_6}, only six virality scores are obtained under Wiener index.
In contrast, cascade virality provides improved ability to
distinguish different types of cascades.
In other words,
not considering the root node in a cascade tree has largely limited the expressive power of Wiener index in capturing the virality of cascades.
It's expected that with the increase of cascade size, 
the ambiguities between cascades will become more and more severe under Wiener index.
However, cascade virality, which considers the direction of the cascade flow, 
will be able to largely avoid these ambiguities compared with Wiener index.
}

The obtained cascade virality has good properties as well.
For totally broadcast cascade (e.g., Fig. \ref{fig:viral_exs}(a)), the virality is minimized and should be 1, as the root node directly connects all other nodes.
For totally viral cascade of size $N$ (e.g., Fig. \ref{fig:viral_exs}(i)), the virality is maximized and should be $(N-1)(N+2)/4$ as the cascade tree is ultimately a path graph with the root node located at one end.
For any other type of cascade, 
% the virality of it locates at the range 
{its virality is within the range}
[1, $(N-1)(N+2)/4$].
It's also anticipated that,
for cascades with a fixed size,
those
with {more} generations (i.e., depths) are generally more viral than their counterparts with smaller generations.
For example, as shown in Fig. \ref{fig:viral_exs}, cascades with a depth of 3 (Fig. \ref{fig:viral_exs} (f)-(h)) are generally more viral than cascades with a depth of 2 (Fig. \ref{fig:viral_exs} (b)-(e)) and {the} cascade with a depth of 1 (Fig. \ref{fig:viral_exs} (a)).

\begin{figure*}[!t]
\centering
\includegraphics[width=\hsize]{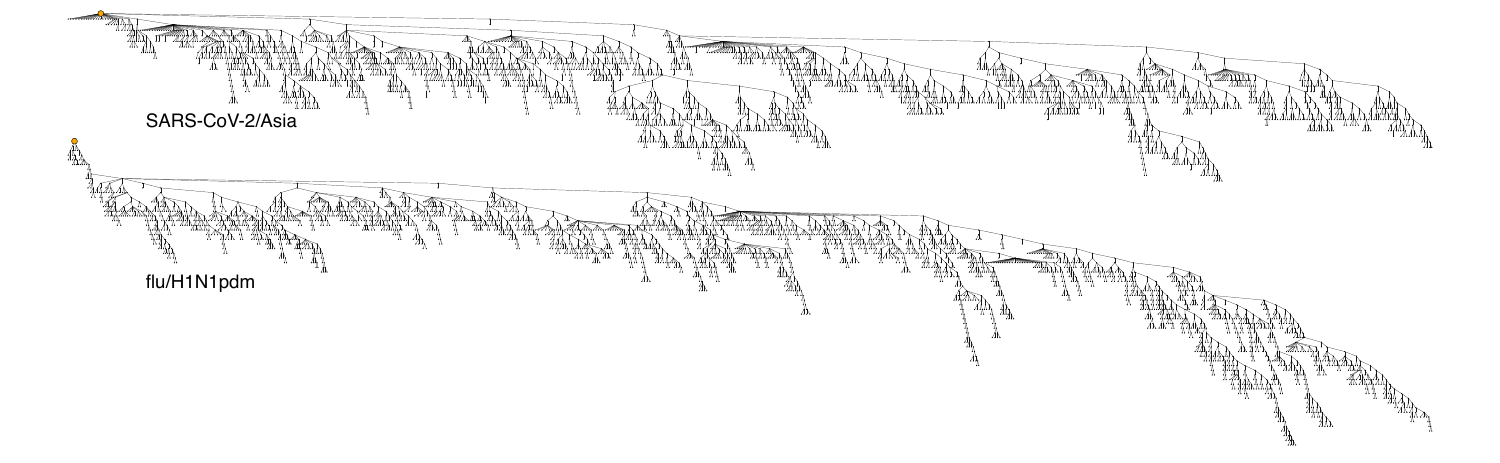}
\caption{\label{fig:virus_strains}
Cascades of virus evolution.
The upper and lower diagrams show the sampled genomic epidemiology of 
% the novel coronavirus COVID-19 pandemic
{SARS-CoV-2} in Asia and the influenza H1N1pdm, respectively.
}
\end{figure*}

\begin{figure*}[!tb]
\includegraphics[width=\textwidth]{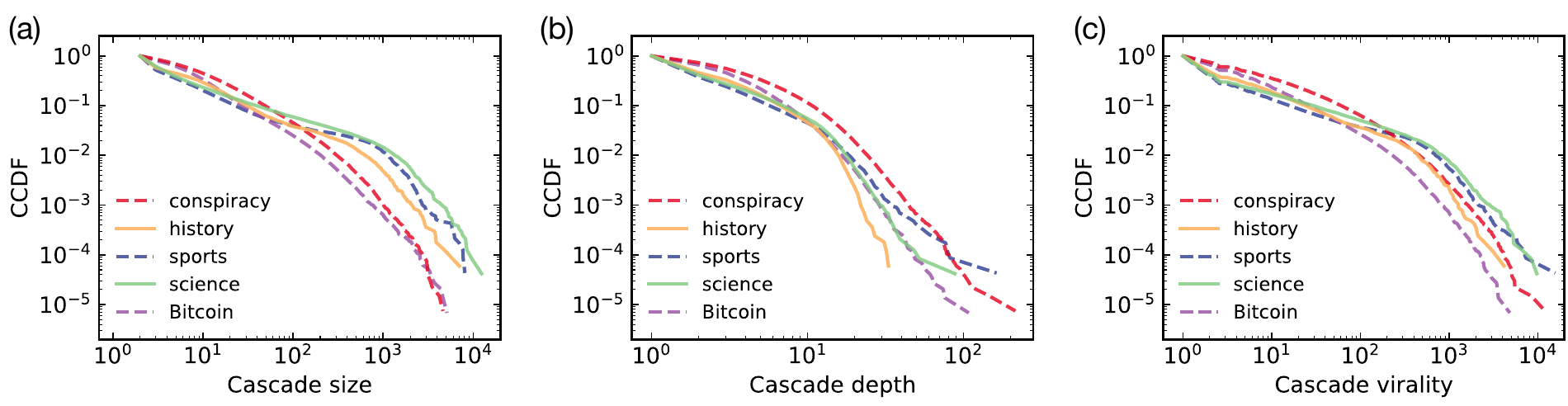}
\caption{\label{fig:cas_analysis}
Analysis of discussion cascades.
(a) CCDF (Complementary Cumulative Distribution Function) of cascade size.
(b) CCDF of cascade depth.
(c) CCDF of cascade virality.
}
\end{figure*}

\begin{table}[!b]
    \centering
    \setlength{\tabcolsep}{6pt}
    \begin{tabular}{llrr}
    % \toprule
    \hline\hline
    \noalign{\smallskip}
                          virus &  size &  depth &  virality \\
    % \midrule
    % \noalign{\smallskip}
    \hline
    \noalign{\smallskip}
    {SARS-CoV-2}/Africa &  2646 &     37 &   2888.97 \\
    {SARS-CoV-2}/Asia &  3431 &     30 &   3635.84 \\
    {SARS-CoV-2}/Europe &  8241 &     51 &   8379.32 \\
    {SARS-CoV-2}/N. America &  6309 &     42 &   6057.99 \\
    {SARS-CoV-2}/Oceania &  3833 &     36 &   3895.62 \\
    {SARS-CoV-2}/S. America &  2696 &     34 &   2847.81 \\
    % \noalign{\smallskip}
    \hline
    % \noalign{\smallskip}
    flu/H1N1pdm &  3769 &     65 &   5021.89 \\
    flu/H3N2 &  3826 &     71 &   6592.46 \\
    flu/Victoria &  3140 &     66 &   3827.23 \\
    flu/Yamagata &  2654 &     48 &   2877.85 \\
    % \bottomrule
    % \hline\hline
    \noalign{\smallskip}\hline\hline
    \end{tabular}
    \caption{Quantifying cascade virality on the phylogeny of virus strains.}
    \label{tab:virus_strains}
\end{table}

\subsection{Application on empirical cascades}

\subsubsection{Virus strains}

We first apply cascade virality on the phylogeny of virus strains.
Fig. \ref{fig:virus_strains} shows the phylogenetic tree of two viruses--one is the ongoing novel coronavirus 
% COVID-19 
{(SARS-CoV-2)}
in the region of Asia (denoted as {SARS-CoV-2}/Asia) and the other is the influenza H1N1pdm (denoted as flu/H1N1pdm)--obtained from Nextstrain \cite{hadfield2018nextstrain}.
% sequenced during their outbreaks.
% 
Note that 
the leaf nodes represent the sampled strains, while the rest of nodes represent inferred common ancestors,
and edges indicate the evolutionary relationships \cite{hadfield2018nextstrain, young2019phase}.

Intuitively, flu/H1N1pdm tends to evolve in a more viral way than {SARS-CoV-2}/Asia.
In fact,
the quantified virality of flu/H1N1pdm cascade (with 3769 nodes and 65 generations) is 5021.89,
while 
the quantified virality of {SARS-CoV-2}/Asia cascade (with 3431 nodes and 30 generations) is 3635.84.
More detailed descriptions of evolutionary cascades of
{SARS-CoV-2} in six continents and four kinds of influenza viruses
are further shown in Table \ref{tab:virus_strains}\footnote{The raw genomic epidemiology data are acquired from and publicly available at \url{https://nextstrain.org}. Accessed: 2020-04-29}.
% 
% As we can see from the table,
Based on the obtained data,
% it appears that 
we find that
flu strains evolve into more generations than currently available {SARS-CoV-2} strains
since three out of four flu cascades go beyond the depth of 60 but no regional cascades of {SARS-CoV-2} do so in current stage.
% 
% It appears that
Except for two large {SARS-CoV-2} cascades in Europe and North America,
the rest of {SARS-CoV-2} cascades 
% are comparable with flu cascades in terms of size but 
% seem
seem less viral than flu cascades.
Nevertheless,
it's foreseeable that the virality of {SARS-CoV-2} cascades would increase dramatically in future 
if the rapid proliferation of {SARS-CoV-2}
is still out of control {around the globe}.

\begin{figure*}[!htb]
\centering
\includegraphics[width=16cm]{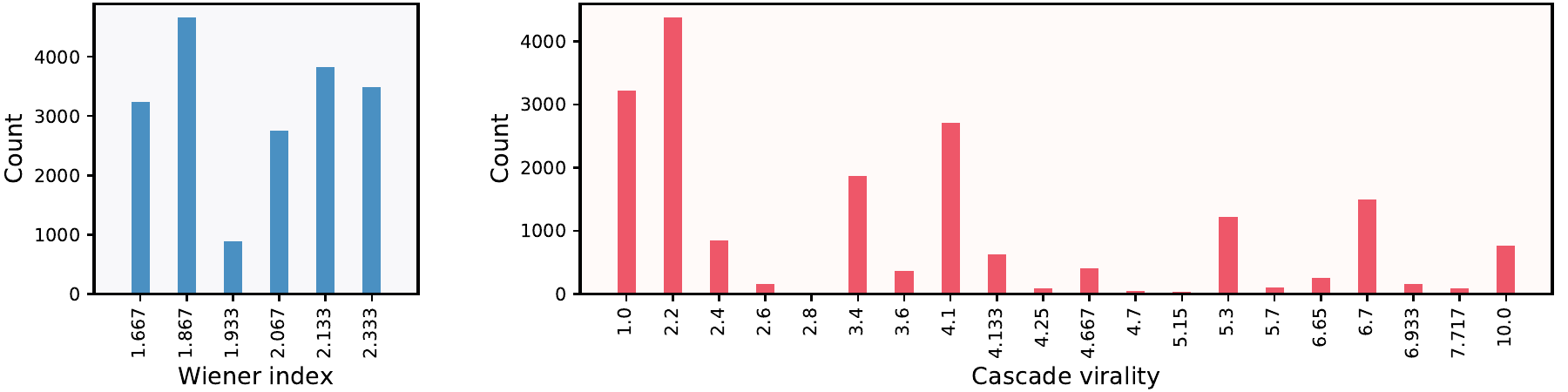}% Here is how to import EPS art
\caption{\label{fig:virality_cas_6}
Distributions of Wiener index and cascade virality for discussion cascades with 6 nodes. There are 18865 such cascades in total from the Reddit discussion data. Each bar corresponds to the count of cascades at specific virality scores under Wiener index or cascade virality.
}
\end{figure*}

\begin{figure*}[!htb]
\centering
\includegraphics[width=16cm]{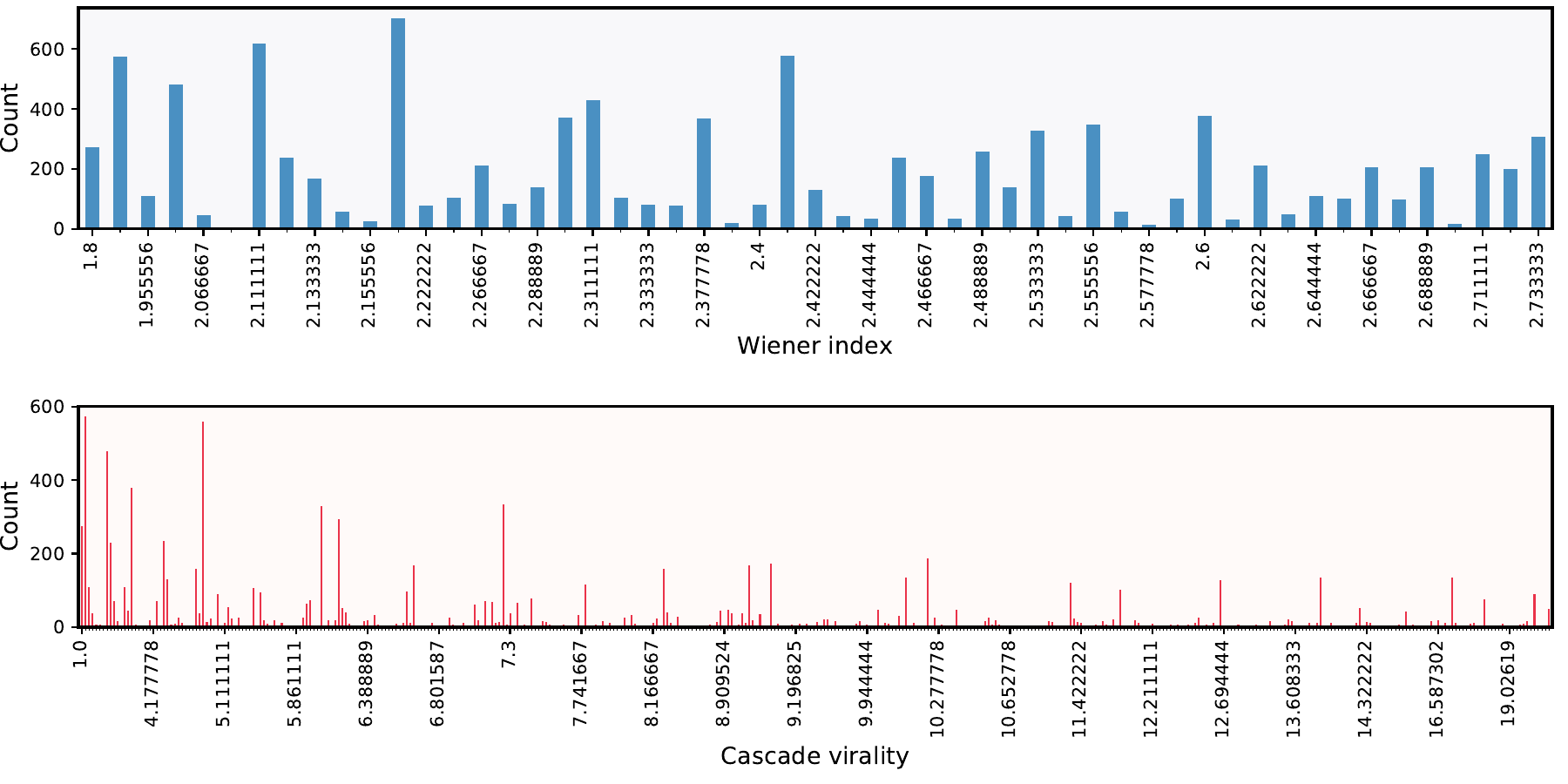}% Here is how to import EPS art
\caption{\label{fig:virality_cas_10}
Distributions of Wiener index and cascade virality for discussion cascades with 10 nodes. There are 10141 such cascades in total from the Reddit discussion data. Each bar corresponds to the count of cascades at specific virality scores under Wiener index or cascade virality.
}
\end{figure*}

\subsubsection{Discussion cascades}

The obtained cascades of virus strains can be considered as very rare events,
we then apply cascade virality on very general cases with large-scale online discussion data as examples.
% rare events
% focus on more general case
% 
A discussion thread can be structurally represented as a rooted cascade tree, where the root node denotes the post itself while each other nodes denotes a comment under this thread \cite{krohn2019modelling}.
Leveraging discussion cascade data from five online communities in Reddit--a social news aggregation and discussion website--we conduct an analysis of cascade virality and its relation with cascade size and depth.
% we conduct a in-depth analysis of cascade .
% we aim to quantify

The dataset includes discussion cascades in five online domains, namely r/conspiracy, r/history, r/sports, r/science and r/Bitcoin,
in the year of 2017 \footnote{The raw data of Reddit are acquired from and publicly available at \url{https://files.pushshift.io/reddit}}.
There are 344 396 cascades in total, including 133 917 in conspiracy, 16 741 in history, 23 178 in sports, 23 871 in science and 146 689 in Bitcoin.
The CCDFs (Complementary Cumulative Distribution Functions) of cascade size, depth and virality are shown in Fig. \ref{fig:cas_analysis}.
As we can see from the figure, most of the cascades have a size of less than 100 (Fig. \ref{fig:cas_analysis}(a)) and a depth of less than 10 (Fig. \ref{fig:cas_analysis}(b)) across domains.
We also note that conspiracy cascades are generally more viral than other kinds of cascades like science and sports (Fig. \ref{fig:cas_analysis}(c)).

\begin{figure*}[!t]
\includegraphics[width=\textwidth]{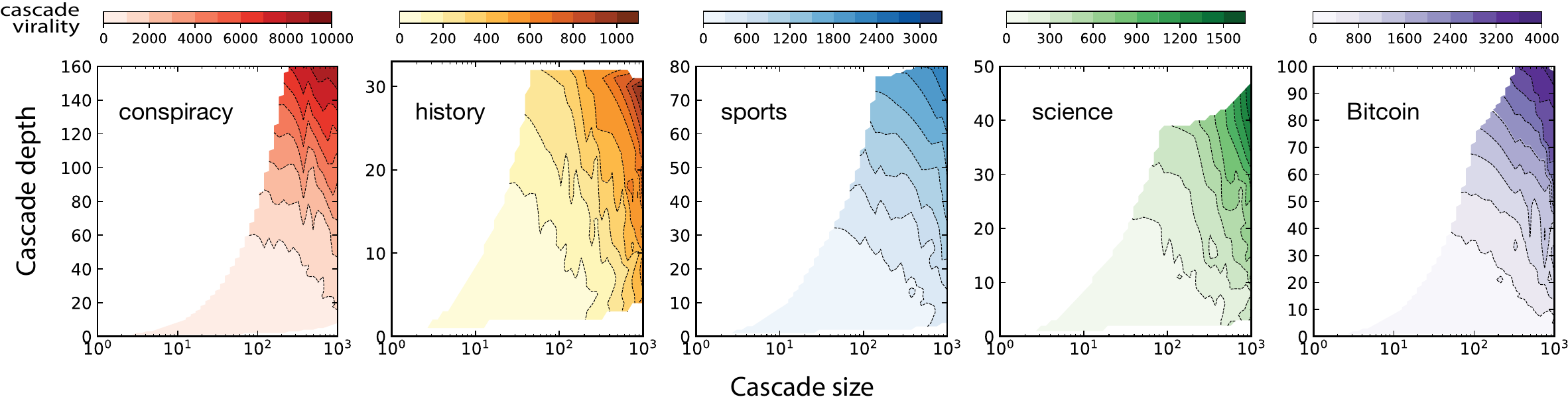}
\caption{\label{fig:virality_contour}
Contour plots of cascade virality versus cascade size and depth.
Color areas in each panel represent cascade virality scores in each domain,
and darker regions indicate higher cascade virality scores.
{
The contours are obtained by interpolating the data to regular grids using unstructured triangular grid functions from \textit{Matplotlib}.
% \footnote{\url{https://matplotlib.org/3.1.1/api/tri_api.html}}.
Blank areas indicate no data available at specific points after the interpolation.
For ease of visualization, contour lines are set as 10 levels for each domain.
}
}
\end{figure*}

\begin{figure}[!t]
\centering
\includegraphics[width=6cm]{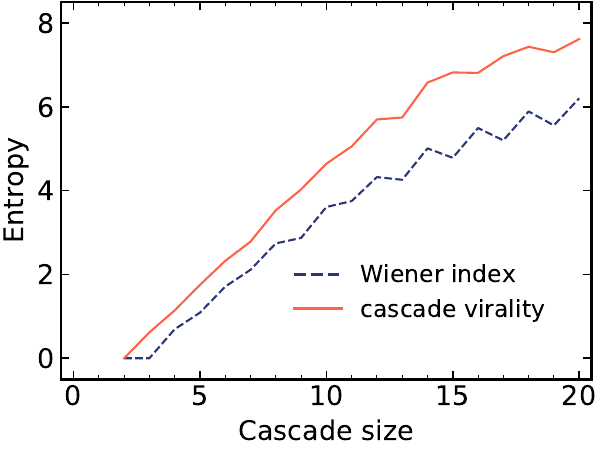}% Here is how to import EPS art
\caption{\label{fig:entropy_analysis}
Entropy analysis based on discussion cascades from Reddit.
}
\end{figure}

Utilizing these empirical discussion cascades, we empirically evaluate the performances of Wiener index and cascade virality in characterizing the virality of cascades.
For cascades with 6 nodes,
Fig. \ref{fig:virality_cas_6} illustrates the number of cascades located at each virality scores under Wiener index (shown in blue) and cascade virality (shown in red).
As shown in the figure, cascades are distributed on six Wiener index scores and twenty cascade virality scores.
The disparities between the distributions of Wiener index and cascade virality become even clear for discussion cascades with 10 nodes (Fig. \ref{fig:virality_cas_10}).
Specifically, Wiener index results in 53 virality scores, but cascade virality is able to quantify 412 levels of virality scores for the same set of cascade data.

To quantify such disparities between Wiener index and cascade virality described above, we conduct an entropy analysis for cascades with equal size.
Given a set of cascades with $N$ nodes, with possible virality scores $v_i$, each with probablity $P({v_i})$, entropy of the virality score distribution is defined as follows:

\begin{equation}
    H_N=-\sum_{i}P({v_i})lnP(v_i).
\end{equation}

% \noindent
For discussion cascades with equal size, we obtain an entropy for Wiener index and another for cascade virality. Fig. \ref{fig:entropy_analysis} shows the comparisons between Wiener index and cascade virality with the discussion cascade size increasing from 2 to 20.
For discussion cascades with 2 nodes, the obtained entropy values via Wiener index and cascade virality show no differences as both of them are equal to 0 in this scenario.
For discussion cascades with 3 nodes, Wiener index will output an unique virality score as it doesn't consider the positions of the root nodes, but cascade virality is able to disambiguate cascades with the root node located at one end and cascades with the root node located at the center (see Fig. \ref{fig:cas_structure} for detailed illustrations).
Therefore, we find a higher entropy of cascade virality than that of Wiener index for cascades with 3 nodes in Fig. \ref{fig:entropy_analysis}.
And more importantly, for increasing cascade size, we find that cascade virality consistently outperforms Wiener index according to the entropy analysis on empirical cascades (Fig. \ref{fig:entropy_analysis}).
Based on the above illustrated cascade examples and analyses,
it's becoming clear that without considering the root nodes, Wiener index is not adequate to characterize the virality of cascades,
as many cascades will be deemed as the same if the root node is neglected.
However, cascade virality, which takes the root node into account, 
can largely disambiguate cascades that Wiener index has failed to do so.

With a further exploration
of empirical discussion data,
we find that 
both cascade size and depth empirically contribute to an increase of cascade virality across domains (Fig. \ref{fig:virality_contour}).
As shown in Fig. \ref{fig:virality_contour},
cascade virality scores become larger and larger
from the lower left to the upper right of the graph in each subplot\footnote{{The contours are obtained using tools from \textit{Matplotlib}: \url{https://matplotlib.org/3.1.1/api/tri_api.html}}}.
In other words, 
for cascades with equal sizes, those whose depths are deeper tend to have higher cascade virality scores.
Likewise, for cascades with equal depths, those whose sizes are larger are also more likely to have higher cascade virality scores.
In summary, in light of the discussion cascades from five online domains,
we find empirical evidence that
cascade virality is positively correlated with
both cascade size and depth.

\section{Modeling cascade growth}

Preferential attachment,
especially for Barab\'{a}si-Albert (BA) preferential attachment, has been widely adopted to model the growth of complex networks \cite{barabasi1999emergence, krapivsky2000connectivity, krapivsky2001organization, golosovsky2018mechanisms, newman2001clustering, jeong2003measuring, eisenberg2003preferential, capocci2006preferential, young2019phase}.
Specifically, it assumes that,
at each time step,
the probability that {an incoming} node added to node $i$ is proportional to node $i$'s degree $k_i$ in the network.
More generally,
one might
consider a non-linear attachment kernel,
where the attachment is proportional to node's degree raised to the power $\gamma$,
% 
% instead of a linear 
% That is to say,
\begin{equation}
    p_{i}=\frac{k_{i}^\gamma}{\sum_{j}k_{j}^\gamma},
\end{equation}
where $k_j$ denotes the degree of node $j$ and the sum is over all nodes already present in the network, and {the} parameter $\gamma$ controls the attachment intensity.

\begin{figure}[t]
\centering
\includegraphics[width=\hsize]{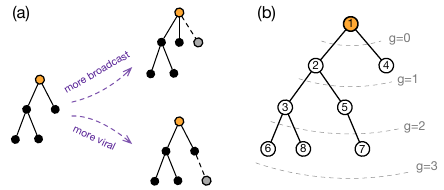}
\caption{\label{fig:broad_vs_virality} 
(a) Schematic of broadcast versus viral growth.
(b) Genealogy of a cascade tree. Root node is shown in color, and dashed lines indicate different genealogy levels of the root node (node 1).
}
\end{figure}

As one special case of complex network, 
% cascade tree 
a tree-structure network
is achieved when each new node only connects to one (instead of multiple) existing node \cite{krapivsky2000connectivity}. 
Recalling the way we quantify cascade virality,
it's anticipated that 
during the growth of a cascade,
new node attachment to long-range leaf nodes would result in a more viral cascade than the attachment to nodes that are close to the root node (Fig. \ref{fig:broad_vs_virality}(a)).
To mimic the growth of cascades and particularly control the virality of cascades during the growth process, 
we propose a genealogy-based preferential attachment mechanism
inspired by classical degree-based attachment kernel.
We will illuminate the proposed model in detail below.

\subsection{Cascade growth based on network genealogy}
For a cascade tree,
once the root node is determined,
{its genealogy}
is readily available.
Note that, for any node in a cascade tree, 
{its genealogy}
can be represented by a (sub)tree rooted at itself.

Fig. \ref{fig:broad_vs_virality}(b) illustrates such an example tree rooted at node 1, where the genealogy levels of node 1 are denoted by $g$.
Specifically,
nodes 2-8 are all descendants of node 1, resulting in a genealogy of size 8 for node 1 (including node 1 itself),
while only nodes 3 and 5-8 are node 2's descendants in the rooted tree, thus resulting in a genealogy of size 6 for node 2 (including node 2 itself).
More importantly,
the genealogy of node 1 at level 1 includes nodes 1, 2 and 4, 
resulting in a genealogy of size 3 at this level;
while the genealogy at level 2 includes two more nodes (nodes 3 and 5), suggesting a genealogy of size 5 at this level.
Similarly, for genealogy of node 2, there are three nodes (nodes 2, 3 and 5) at level 1
and six nodes (nodes 2, 3 and 5-8) at level 2.

Formally, for node $i$ in a cascade tree, 
{its complete genealogy}
contains itself and all the descendants of it,
but the genealogy at level $g$ contains itself and its descendants within $g$ generations only.
For the sake of consistency, the complete genealogy is also denoted as genealogy at level \textit{max}.

Here, we propose a genealogy-based preferential attachment model,
where new node attachment relies on a node's genealogy size at specific genealogy levels instead of degree.
Namely,

{
\begin{equation}
\label{eq:genealogy}
    p_{i}=\frac{d_{ig}^\gamma}{\sum_{j} d_{jg}^\gamma},
\end{equation}
where $d_{jg}$ is the genealogy size of node $j$ at level $g$, and {the} parameter $\gamma$ controls the attachment intensity.
}

Clearly,
$\gamma=1$ recovers the linear attachment kernel
while $\gamma=0$ recovers the uniform attachment kernel.
When
$\gamma>1$, the attachment kernel becomes super-linear,
while in the interval
$0<\gamma<1$, the attachment kernel becomes sub-linear.
For $\gamma<0$,
the model 
favors homogeneous attachment
where {incoming} nodes are preferentially attached to existing nodes with small genealogy sizes.

It's anticipated that
in the limit $\gamma\rightarrow+\infty$, 
the model would
generate totally broadcast cascade {trees}, where the root node directly connects all other nodes.
In the limit $\gamma\rightarrow-\infty$, 
the model would
generate totally viral cascade {trees}, where each node has one successor only except for the last node which acts as the leaf node.

\begin{figure}[t]
\centering
\includegraphics[width=\hsize]{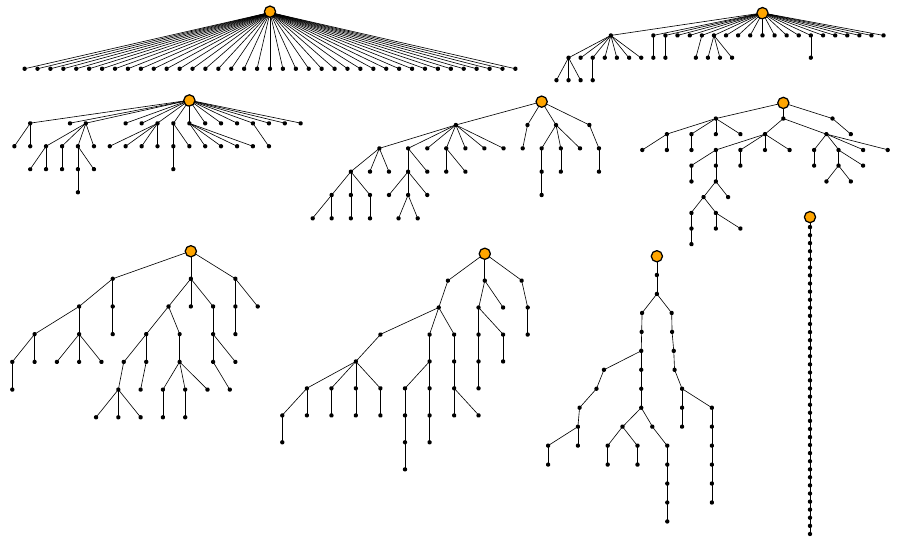}
\caption{\label{fig:cascades_genealogy_max}
{
Synthetic cascades when the complete genealogy is considered in the attachment kernel
with $\gamma\in\{5.0, 1.5, 0.75, 0.25, 0, -0.25, -1.0, -2.5, -10.0\}$ (from top left to bottom right).
Taking cascades of size 40 for illustration purposes only.
}
}
\end{figure}

% \section{Model Fiting}
% \section{Results}

\subsection{Numerical results on synthetic cascades}

We numerically evaluate the effect of parameter $\gamma$ in driving the virality of cascades.
% 
% We show 
Fig. \ref{fig:cascades_genealogy_max} shows
the generated cascades based on complete genealogy
with the varying 
% $\gamma\in\{5.0, 1.5, 1.0, 0.5, 0.2, 0, -0.2, -2.5, -10.0\}$
$\gamma\in\{5.0, 1.5, 0.75, 0.25, 0, -0.25, -1.0, -2.5, -10.0\}$
when the cascade size is set as 40.
As shown in the figure,
for $\gamma>1$,
the generated cascades tend to condense on nodes at low depths (particularly on the root node),
where the number of successors diminishes with the increase of distance from the root node.
When $\gamma$ is set by large positive values, such as 5.0 in the example, the generated cascade will condense completely on the root node, which corresponds to a totally broadcast cascade.
With the decrease of $\gamma$, the condensation state diminishes gradually
and even becomes trivial for 
negative values of $\gamma$.
When $\gamma$ is set by negative values with large magnitude, such as -10 in the given example, the generated cascade will form a long path with the root node located at one end, which corresponds to a totally viral cascade.
% 

% We further 
{Further, we}
numerically investigate the relationship between cascade virality and parameter $\gamma$ in different cascade growth models.
We consider not only the complete genealogy size but also genealogy at level 1 and classic degree based attachment kernel.
Specifically,
cascade size is set as 100,
$\gamma$ is set in the range of [-15, 10] with a step of 0.1, 
and 100 cascades are generated under parameter $\gamma$ in each growth model.

\begin{figure}[t]
\centering
\includegraphics[width=7cm]{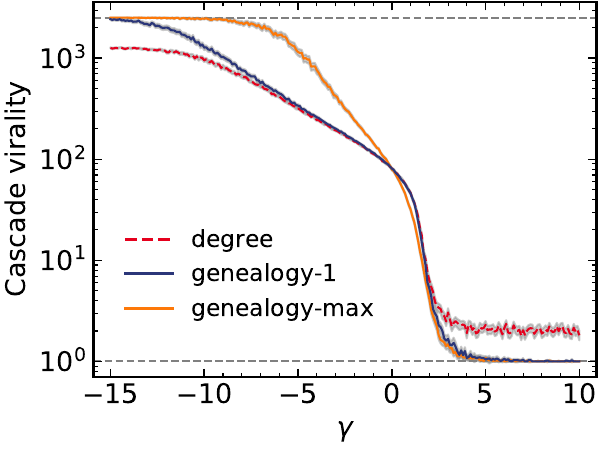}
\caption{\label{fig:virality_vs_gamma} 
Cascade virality versus parameter $\gamma$ according to different growth mechanisms (using cascades of size 100 for illustration).
Cascade virality bounds are shown in grey dashed horizontal lines.
Shading areas (in grey) indicate bootstrapped 95\% confidence intervals obtained from 100 random simulations under each parameter $\gamma$.
As shown in the figure, classic degree-based mechanism cannot efficiently exhaust the complete virality space.
}
\end{figure}

In Fig. \ref{fig:virality_vs_gamma}, we show the cascade virality 
% with the increase of 
{for increasing}
$\gamma$.
Note that, for cascades of size 100, their virality scores are located 
% at 
{within}
the range [1, 2524.5].
The upper and lower bounds are shown in grey dashed horizontal lines in the figure.
Clearly, 
cascade virality decreases with the increase of parameter $\gamma$ in each model,
but classic degree-based kernel cannot efficiently exhaust the complete virality space.
With the increase of $\gamma$,
two genealogy-based models 
ultimately
converge to completely condensate state,
which is not observed for degree-based kernel.
In summary, 
genealogy-based models provide much more strong representation power than classic degree-based model in generating cascades with varying virality scores.

\begin{figure*}[tb]
% \centering
\includegraphics[scale=0.53]{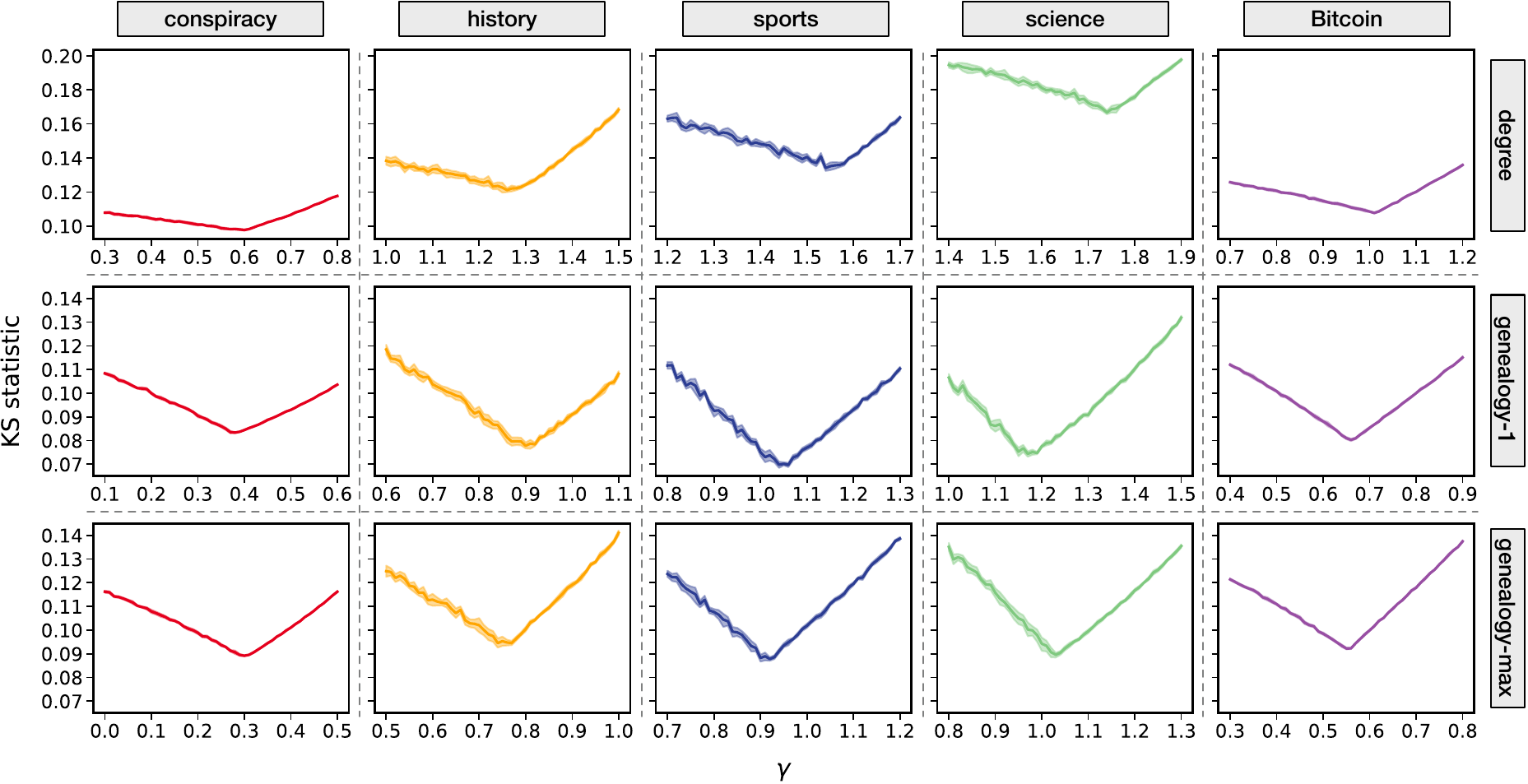}
\caption{\label{fig:KS_gamma_combined}
KS-statistic versus parameter $\gamma$ in numerical simulations.
KS-statistic measures the maximum  discrepancy  between  
the cumulative  distribution functions (CDFs) of two virality distributions:
the empirical virality distribution and the generated virality distribution. 
Shading areas indicate bootstrapped 95\% confidence intervals obtained from ten random simulations.
}
\end{figure*}

\subsection{Numerical results on empirical cascades}

We test the performance of cascade growth models on empirical discussion cascades described above.
For each discussion cascade, a synthetic cascade with the same size is generated according to each growth model.
We then compare the virality of generated cascades with that of real cascades in terms of the distribution of virality scores.
The comparison can be done by Kolmogorov-Smirnov (KS) statistic between the generated cascade virality scores and real cascade virality scores.

The KS statistic of two distributions is given by the 
maximum discrepancy between their cumulative
distribution functions (CDFs),
and has been applied in 
assessing the goodness of fit between two degree distributions \cite{clauset2009power, young2019phase}.
Generally,
smaller KS statistic favors better fit.
Given the empirical virality distribution,
we estimate $\gamma$ by minimizing the KS statistic averaged over ten random simulations under each parameter $\gamma$.
The optimal $\gamma$ for each model under each community is chosen as the $\gamma$ that results in the minimum KS statistic over the parameter space.

% In practice, 
% 
For ease of computation, 
cascades whose sizes are larger than 100 or the 95th percentile, whichever is greater, are omitted in the analysis.
Cascades of size 2 are also not included as any cascade growth model can exactly generate the same kind of cascade with 2 nodes.
Fig. \ref{fig:KS_gamma_combined} presents the results for degree, genealogy-1 and genealogy-max based models,
where smaller minimum KS statistic indicates better model under each community.

As shown in the figure,
the minimum KS statistics achieved by
degree-based model 
are larger than 0.097 across all five domains,
while genealogy-based models show better performance 
with the minimum KS statistics all less than 0.096 (see also Table \ref{tab:minimum_KS} for details).
Specifically, genealogy-1 achieves better performance than genealogy-max.
In summary,
leveraging data of discussion cascades from five online domains,
we demonstrate the performance of
genealogy-based models to mimic the growth of cascades.

\begin{table}[!htb]
    \centering
    \caption{The minimum KS statistic for each model.}
    \label{tab:minimum_KS}
    \setlength{\tabcolsep}{2.5pt}
    \begin{tabular}{llllll}
        \hline\hline
                & conspiracy  &  history &  sports & science & Bitcoin \\
        \hline
        degree    &       0.098 &    0.121 &   0.134 &    0.167 &    0.108 \\
        genealogy-1 &       0.083 &    0.078 &   0.070 &    0.074 &    0.080 \\
        genealogy-max      &       0.089 &    0.095 &   0.088 &    0.090 &    0.092 \\
        \hline\hline
    \end{tabular}
\end{table}

\section{Conclusion}

In this work,
we introduce a root-aware approach to {quantifying} the virality of cascades.
The proposed approach considers cascades as directed instead of undirected trees,
{thus alleviating the effects induced by graph isomorphism that previous studies have overlooked}.
Using synthetic and empirical cascades,
we show the property and utility of the proposed approach
and particularly
the relationships between cascade virality and cascade size and depth.
In doing so, we find that both cascade size and depth empirically contribute to an increase of cascade virality from the obtained large-scale discussion data.

To mimic the growth of cascades
as an interpolation between broadcast and viral proliferation,
we further introduce a cascade growth model.
Specifically,
the attachment kernel is constructed
based on node genealogy rather than node degree.
With the decrease of
the attachment intensity,
the generated cascade grows from a broadcast to a viral way.
Numerical simulations on synthetic and empirical cascades reveal the advantage of genealogy-based attachment kernel over degree-based attachment kernel
in characterizing the virality of cascades.

Our work contributes to a further {understanding} of how to quantify the virality of cascades and 
provides a rich avenue for compelling applications.
For example, as we have demonstrated in the main text, conspiracy and science cascades display quite different patterns in the distribution of cascade virality.
As such, the proposed approach to {quantifying} cascade virality may provide 
{improved ability}
to distinguish different types of cascades.
Moreover, as shown in the paper, the attachment kernel significantly affects the growth and virality of cascades, which may offer valuable insights to the implementation of viral marketing in practice.
In this work,
we mainly address tree-structure cascades,
but future works may consider
more complex
diffusion structures,
such as the locally tree-like diffusion process which is prevalent in network spreading.
We have applied the proposed measure to capture the virality of {SARS-CoV-2} and influenza strains,
but
future works can directly apply it on infection data
(once such data are available)
to have a better {understanding} of the spread of virus.

% \newpage

\vspace{30mm}

\appendix

% \section{Appendixes}
\section{Property}

\textit{
For a cascade $T$ of size $N (N\geq 2)$, 
the cascade virality of it satisfies the following:
% $1\leq V \leq \frac{(N-1)(N+2)}{4}$
$1\leq V_T \leq (N-1)(N+2)/4$.
}

% \newpage

% \paragraph{Syntax}
\vspace{3mm}

\begin{enumerate}

\item Extreme case when the cascade virality is minimized:
% (e.g., FIG. \ref{fig:cas_exs} $(a)$ and $(c)$):
a star graph with the root located at the hub (i.e., totally broadcast),
and the corresponding virality of the cascade is $V_T=1$;

\item Extreme case when the cascade virality is maximized:
% (e.g., FIG. \ref{fig:cas_exs} $(f)$):
a path graph with the root located at one end (i.e., totally viral)

\begin{itemize}

% \item For a complete path cascade of size 1, the cascade virality is 0; 

\item For a complete path cascade $T$ of size 2, the cascade virality is 1;

% For cascade of size 3, the cascade virality is $1+\frac{1+2}{2}=2.5$;

\item For a complete path cascade $T$ of size $N (N\geq 3)$, the cascade virality is 
$V_T=1+\frac{1+2}{2}+...+\frac{1+2+...+(N-1)}{N-1}
=1+\frac{3}{2}+...+\frac{N}{2}
=\frac{2+3+...+N}{2}
=\frac{(N-1)(N+2)}{4}
$;

Taken together, we know that
for a complete path cascade $T$ of size $N$, its cascade virality is
$V_T=\frac{(N-1)(N+2)}{4}$.

\end{itemize}

Therefore, for a cascade $T$ of size $N (N\geq 2)$, 
the cascade virality of it
should lie between $1$ and $(N-1)(N+2)/4$.

\end{enumerate}

% \vspace{10mm}

\bibliography{main} % Produces the bibliography via BibTeX.

\end{document}